## Designing Co-operation in Systems of Hierarchical, Multi-objective Schedulers for Stream Processing

Animesh Dangwal, Yufeng Jiang, Charlie Arnold, Jun Fan, Mohamed Bassem, Aish Rajagopal
Meta Platforms

### 1. Abstract:

Stream processing is a computing paradigm that supports real-time data processing for a wide variety of applications. At Meta, it's used across the company for various tasks such as deriving product insights, providing and improving user services, and enabling AI at scale for our ever-growing user base.

Meta's current stream processing framework supports processing TerraBytes(TBs) of data in mere seconds. This is enabled by our efficient schedulers and multi-layered infrastructure, which allocate workloads across various compute resources, working together in hierarchies across various parts of the infrastructure. But with the ever growing complexity of applications, and user needs, areas of the infrastructure that previously required minimal load balancing, now must be made more robust and proactive to application load.

In our work we explore how to build and design such a system that focuses on load balancing over key compute resources and properties of these applications. We also showcase how to integrate new schedulers into the hierarchy of the existing ones, allowing multiple schedulers to work together and perform load balancing, at their infrastructure level, effectively.

### 2. Introduction:

Meta's stream processing platform manages TBs of data per second. To provide better management, the platform splits application workloads into multiple sets of clusters called tiers[1]. Each tier handles a subset of the workload. As applications can independently expand in resources consumed, this setup poses a challenge to balance the load between tiers. Typically this balancing was done manually, which takes quite a bit of software engineer effort in terms of log searches, and manual data monitoring checks.

This manual method also fails to account for other schedulers which perform their load-balancing duties at lower-level abstractions to tiers (e.g. regions, or hosts)[2, 4], which could result in one or more of these schedulers failing to assign an application optimally, for example, moving an application from tier 1 whose preference is to region A to be closer to its data source, to tier 2, which does not have machines in region A, resulting in high-network costs for the application. With applications[3] sensitive to millisecond delays this could be extremely detrimental.

This current under-defined procedure incurs large overhead, and this paper aims to not only automate this balancing, but ensure app movement decisions are efficient, and work with the underlying system of schedulers, while avoiding unintentional human error using Meta's Rebalancer constraint solver[2].

We identify the following properties to load balance over:

- Task count
- Cpu Utilization
- Memory Utilization

We also ensure additional properties of applications are not violated, such as:

- Critical apps moved less frequently to reduce their downtime caused by app movement.
- Apps moved in accordance to their Service Level Objective (SLO)
- App movements are compatible with lower-level schedulers, such as region based placements to ensure network costs incurred by an application is low.

We also compare our results against a baseline greedy scheduler to validate our solution quality and showcase the need for a holistic view of load balancing rather than greedily moving apps from higher utilized tiers to lower utilized tiers, per resource objective.

Our contributions are:

- Designing and building the StremProcessing Tier Load Balancing(SPTLB) scheduler to eliminate manual intervention when load balancing applications across tiers.
- Establishing the need for the SPTLB scheduler when compared to the baseline greedy scheduler
- Ensuring the SPTLB scheduler fits within the hierarchy of existing schedulers in Stream Processing.

### 3. SPTLB architecture

Manual decisions are made which aren't necessarily optimal or error-free, hence we break down the problem into three stages, explored in the architecture below, as seen in Figure 1.

#### 3.1 Data Collection:

To first understand what we need to load balance, we need to collect data on resource utilization (cpu, memory) and task counts for an app in the tier.

- We use our internal app metadata store to get running apps and their information on SLO and criticality as scores.
- The metadata store also gives us resource monitoring endpoint information per app.
- This endpoint is then used to collect live cpu, memory and task count information.

- We focus on collecting peak resource utilization (99th percentile) when load balancing to account for application scaling during execution
- We additionally collect tier metrics as well in terms of their limits and ideal resource utilization conditions. This helps the scheduler to constrain movements across tiers via these upper limits.

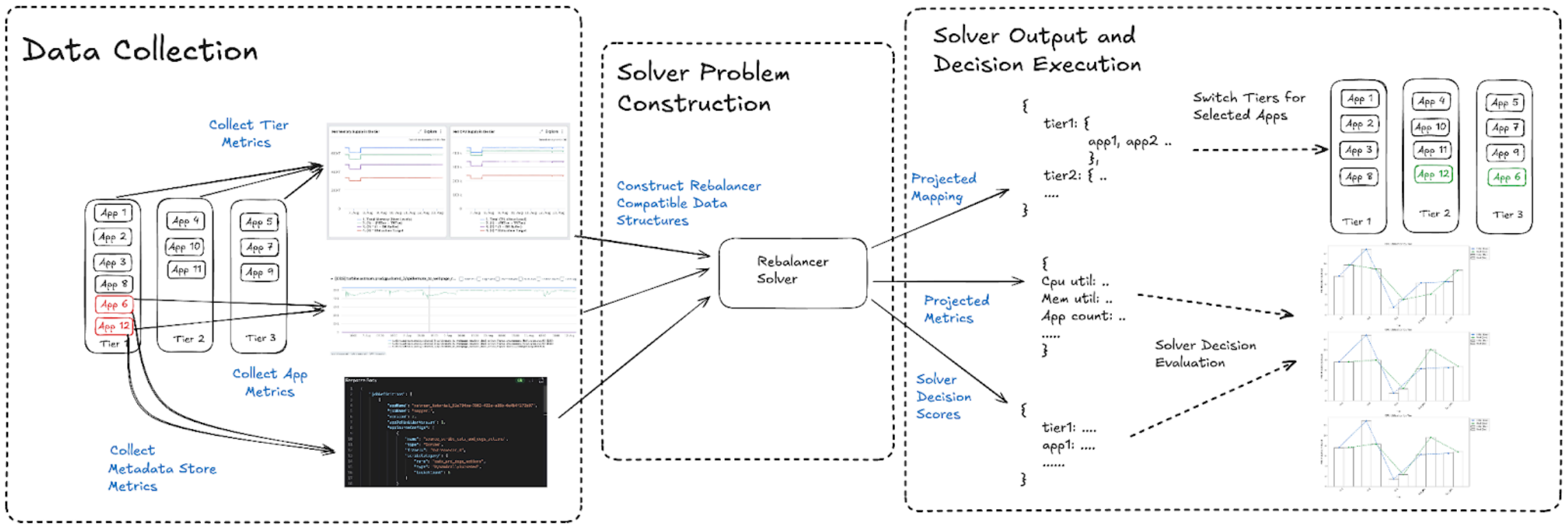

Figure 1: We have the full system (shows example data and not real application information) workflow here with data collection as our first step. The data collected is then converted into Rebalancer compatible data structures after which we use these structures in the final step of tracking projected metrics and evaluating solution quality.

## 3.2 Solver Problem Construction

We now need to define these metrics collected as part of the problem specification for Rebalancer.

There are two halves to constructing the problem for Rebalancer:

- Constructing compliant data structures for the solver to understand the system and its properties
- Modelling the load balancing problem via constraints and goals.

### 3.2.1. Modelling the problem

We focus on modelling the following statements:

**Constraints** (all equally important to be satisfiable to get a valid solution)

1. "Tiers should not exceed their capacity metric for any resource"
    a. This is modelled as a constraint by design as the dimensions on the tier are defined as the headroom capacity, so no solution can ever exceed this for cpu, mem used.
2. "Tiers should not exceed their task count"
    a. Also modelled as a constraint by design as the dimensions on the tier as defined as the task limit for the tier.
3. "App movement must be limited per solution generated by load balancer"
    a. Modelled as a constraint explicitly, minimize movement constraint specifies allowance as x% of total applications across all tiers.
4. "Apps with SLO scores must be in tiers that support said SLO score"
    a. This is modelled as a constraint explicitly, by adding SLO scores as an avoid movement to tiers that do not match said SLO score. This prevents any SLOs from being violated by our SPTLB.

**Goals** (ordered by default priority, all goals always lower priority to constraints)

5. "Tiers resource utilization is preferred to be under utilization limit"
    a. Modelled as a goal to optimize, so valid solutions can violate this, allowing for solutions to be provided when multiple tiers under heavy load
6. "Resource usage is balanced across tiers"
    a. This is modelled as a goal using the resource dimensions (cpu, mem) for apps and tiers. This is relative to each tier, due to the above statements 1, 4
7. "Task count is balanced across tiers"
    a. This is modelled as a goal by balancing apps across tiers. This is also relative and due to the above statements 2, 3
8. "App downtime is low during switch tier"
    a. This is modelled by adding task_count as the cost of movement, so it deters the solver from moving apps with large task counts as much as possible, to reduce downtime for switching tiers

9. “Apps with high criticality scores are not moved frequently”
    a. This is modelled as a goal by adding criticality scores as an affinity for a tier container. It deters the solver from suggesting solutions that move apps with high criticality scores, as decided by the solver on what “high” is relative to the scores of other applications.

We also have the option of solver type in Rebalancer:
- LocalSearch: Greedy exploration of search space to find a solution, can get stuck in local minimums
- OptimalSearch: Provides a linear programming solver to search for optimal/close-to-optimal solutions for the problem, this is usually both the most time consuming solver and the best performing solver in terms of solution quality

We explore these tunings in the results section, Figure 3. While we do have other tuning options possible for SPTLB depending on the prioritization of the goals, the explored results do not provide any significant improvements from the default priorities and hence only use the default priority for all comparisons in Section 4.2.

### 3.3 Solver Output and Decision Execution

We give the output as the projected mappings from tier to app after load balancing and the projected metrics of cpu, memory, app_count/task_count.
- These are used to suggest and give recommendations regarding what apps to move to balance the tiers appropriately.
- The solver decisions can also be evaluated against the greedy scheduler and or potentially human decisions to compare which decision is performant.
- This decision evaluation can also result in finding bugs with the solver in terms of how the tuning knobs/goals and constraints are defined and if they’re followed correctly.
- These are also emitted as metrics in the resource endpoint of the SPTLB.

### 3.4 Hierarchy Integration

The simplest way to introduce cooperation within the system of schedulers without additionally adding complexity is to allow for suggested app-to-tier mappings to pass through to the lower-level schedulers for region and host, as seen in Figure 2.

This way the complexity and time taken to generate a schedule by SPTLB is minimally affected. Alternatively we could introduce more complexity to the SPTLB scheduler and allow it to account for region disparities between the tiers but this leads to not only inefficient solutions but also adds time taken to generate a solution by the SPTLB scheduler. More of this is explored in the results section in Figure 5.

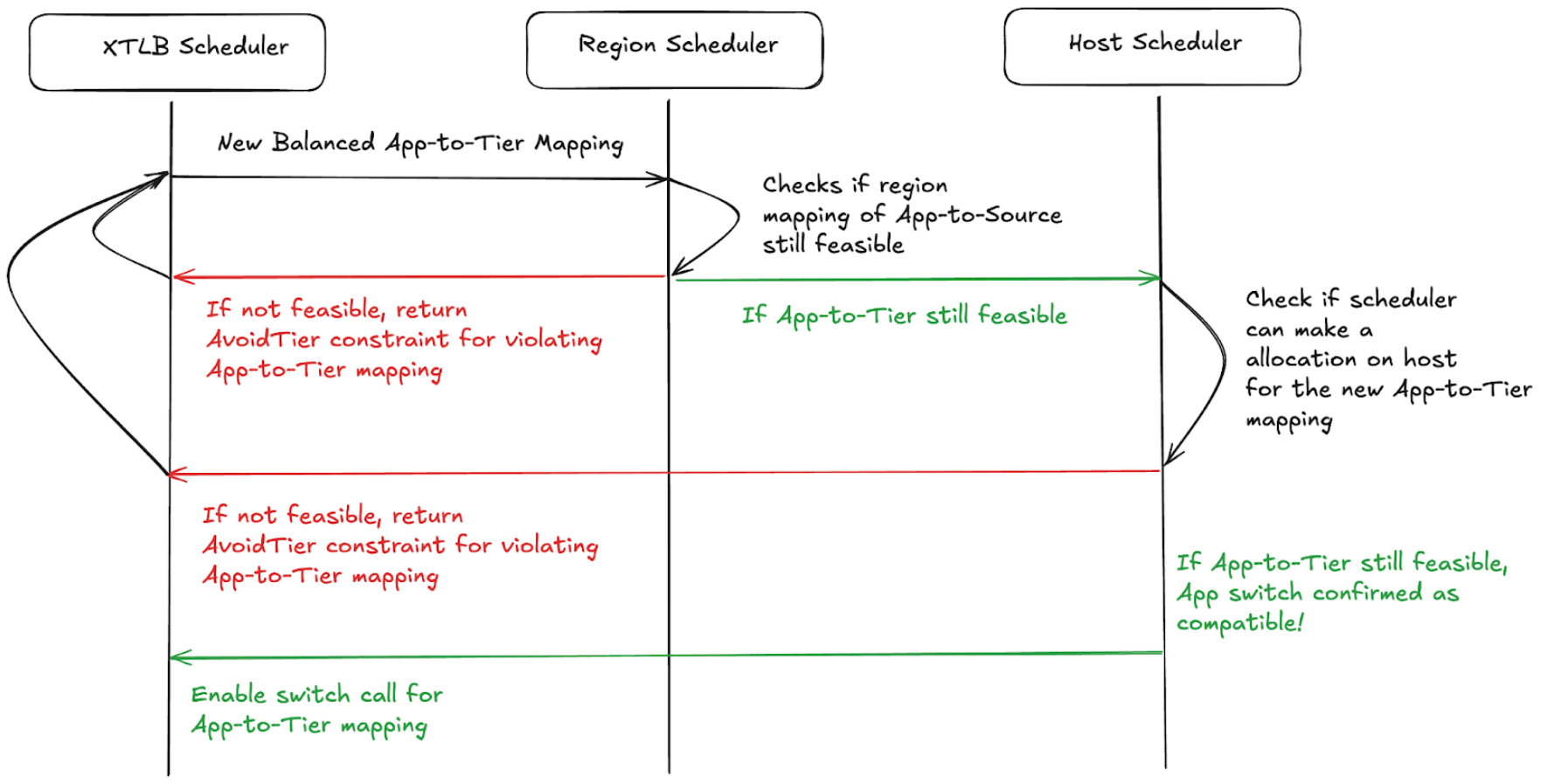

Figure 2: We have the full workflow diagram of the interactions between the schedulers SPTLB with region preference and host allocation. A mapping of apps to tiers is presented to the region scheduler. If it isn’t possible to keep an app near its data source with the given tier, it returns false to the SPTLB scheduler which adds additional avoid constraints, similar to Constraint 3 in section 3.2.1. If the mapping is possible it goes to the next lower-level scheduler the host scheduler, where if there are available hosts to allocate the application to, it accepts the mapping and sends an acknowledgement to SPTLB, however if it fails, similar to before, it returns false to SPTLB which will add an avoid constraint again and resolve the new mapping. These iterations continue until SPTLB times out or the number of iterations limit is reached.

## 4. Solver Algorithm Exploration:

We now compare how these different tuning knobs, such as, local search, optimal search and various timeouts like 30s, 60s, 10minutes, 30minutes, and observe how they affect solution quality, when compared against the simple greedy scheduler.

Our experiment setup is scheduling over live tier data from Meta's clusters where we have 5 tiers, belonging to the following SLO mappings:

- SLO1: tier 1, 2, 3
- SLO2: tier 1, 2, 3
- SLO3: tier 1, 2, 3, 4, 5
- SLO4: tier 4, 5

We also run our greedy scheduler over the same values and present their performance below in Figures 3. We also showcase an analysis of network performance, approximated to the closest millisecond(ms) to highlight which hierarchy integration works better (as seen in Figure 4, 5).

### 4.1 A simple greedy solver

To sanity check our SPTLB scheduler we also compare our results against a basic greedy scheduler.

We designed the greedy scheduler as a stand in for manual decision making where we:

1. Identify the tier with the most resources used given the utilization target (resources used/util target) and least resources used given the utilization
2. Identify the largest app in terms of resources utilized (either cpu, mem, task count) and hasn't already been moved yet
3. Move app to tier with lowest utilization
4. Loop from step 1 until x% of apps moved or timeout reached

We compare against the metrics of task count, mem utilization, cpu utilization in the final tier-app assignments generated by both load balancers.

### 4.2 Results

We present the results in three parts:

1. Resource balancing against greedy schedulers. This will showcase that the SPTLB is effectively performing its load balancing duties
2. Network cost comparison against different hierarchy integrations between SPTLB and other lower-level schedulers
3. Overall performance of the scheduler integration methodologies

#### 4.2.1 SPTLB vs Greedy Scheduler

The graph in Figure 3(a) represents the cpu utilization across 5 of the real tiers used at Meta. We run each scheduler with a timeout of 30s and bound app movement by 10%, therefore only 10% of the total apps across the tiers can be moved during a single call to SPTLB or Greedy Scheduler. We do not showcase optimal search or other timeouts as there is no significant difference in the patterns that emerge in Figure 3(a), (b), (c).

The cpu utilization in Figure 3(a), represented by the y-axis, is a percentage of the total capacity in each tier, therefore the bars are relative to their max capacity of the tiers, represented by the x-axis. The tiers have a default ideal utilization of 70% which is indicated by the orange dashed line. The red line indicates full (100%) capacity for the resource.

The red bars are the initial value of the cpu utilization when the SPTLB collected the data, and the neon-green bar is the final state as load balanced by the SPTLB scheduler. We can see that there is a clear improvement in terms of balancing the utilizations as all cpu values are much closer to each other than the initial state, seen most clearly from tier 3's initial to final utilization. We can also see the variations of the greedy scheduler. Each variation is greedily scheduling the algorithm mentioned in section 4.1 for the particular resource objective, cpu, mem and task count. The greedy-cpu(dark-yellow) is the only greedy variation that performs similarly to the SPTLB solver, while the other variations greedy-mem (blue), greedy-task-count (purple) remain unbalanced. This is the common pattern we see across the other variation in Figure 3 (b), (c) where memory and task count are balanced well by the same mappings SPTLB picked for the cpu graph, however the greedy variants do not perform well across all resource utilizations, effectively always being unbalanced. In fact it even goes as far as exceeding the ideal resource limit for memory utilization graphs in Figure 3(b). This conclusively shows that SPTLB is effective at load balancing across multiple objectives, while maintaining its constraints and objectives.

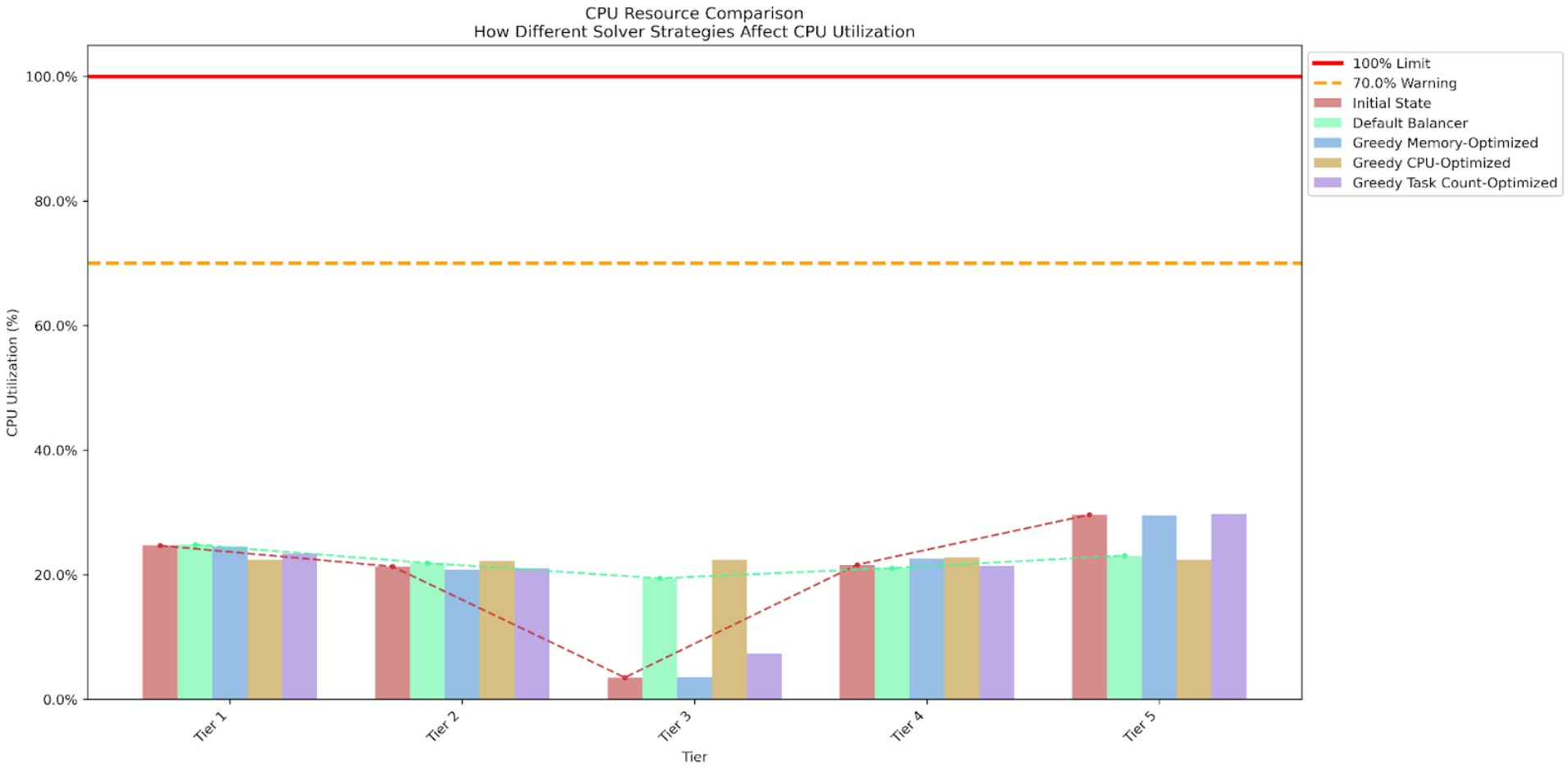

(a) In Figure 3(a) we have compute utilization on the y-axis and the respective tier bar plots in the x-axis. The tiers have a default ideal utilization of 70% which is indicated by the orange dashed line. The red line indicates full (100%) capacity for the resource. The percentage is relative to each tier's capacity limit. The graph represents live data collected from Meta's infrastructure. The red bar represents the initial state before the SPTLB is run. Neon-green represents SPTLB's balanced mapping for app to tier, we can see the bar's are much more comparable in height from the initial red positions. The blue bar represents the greedy scheduler when it prioritizes balancing memory. This is shown here as it clearly leaves cpu unbalanced, while balances memory in Figure 3(b). The dark-yellow bar represents the greedy scheduler that prioritizes cpu balancing which results in comparable values to the neon-green bar, thus showing the baseline greedy solver does perform load balancing effectively enough on one resource at a time, but fails to balance resources across the multiple objectives (cpu, mem, task count). The purple bar represents the greedy scheduler that prioritizes task count, and much like the blue bar of greedy-memory prioritization, it also fails to balance cpu but effectively balances task count in Figure 3(c).

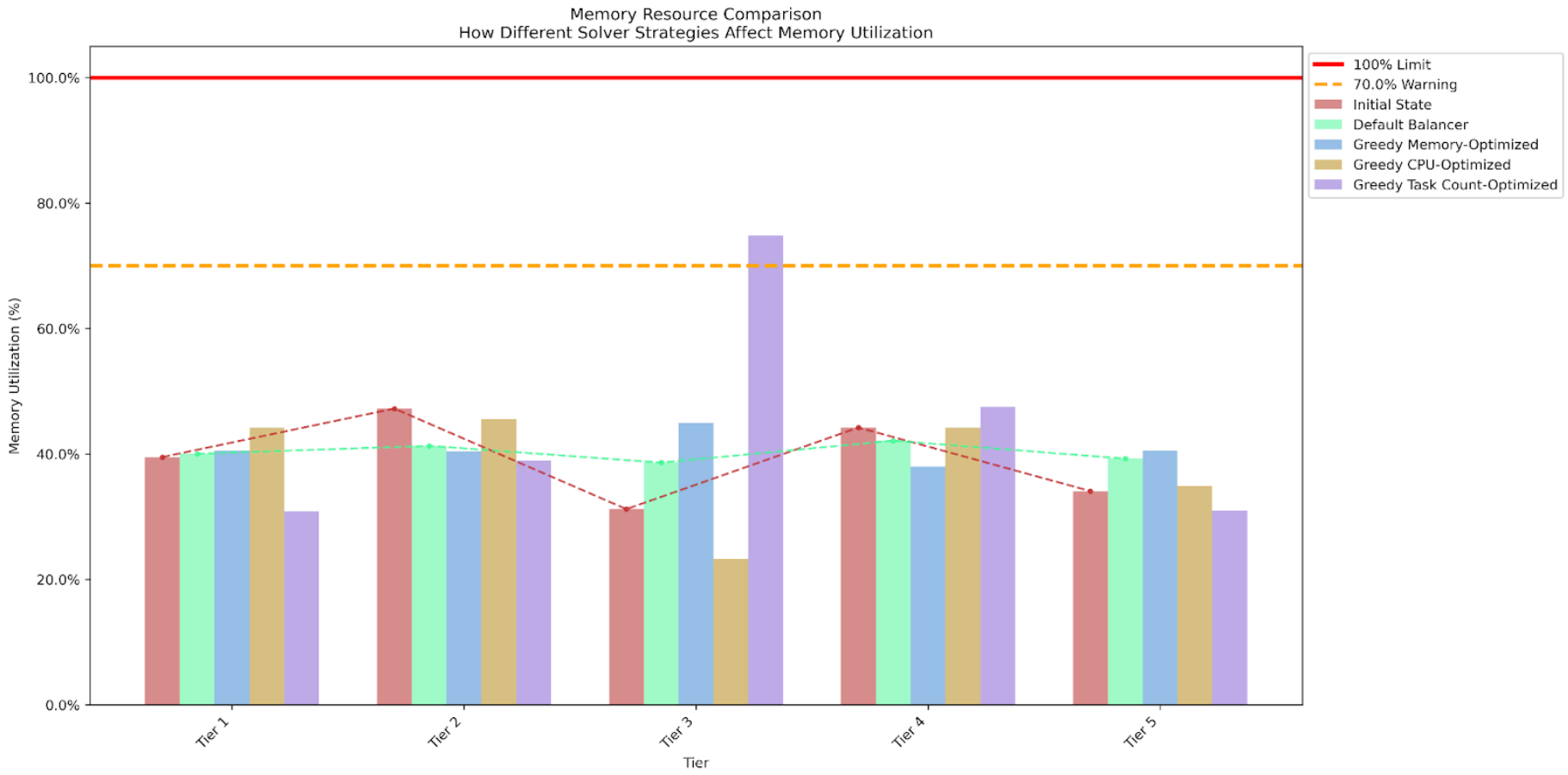

(b) In Figure 3(b) we have memory utilization on the y-axis and the respective tier bar plots in the x-axis. The percentage is relative to each tier's capacity limit. The tiers have a default ideal utilization of 70% which is indicated by the orange dashed line. The red line indicates full (100%) capacity for the resource. We see the same pattern from Figure 3(a), where the baseline greedy solver does perform load balancing effectively enough on one resource at a time, but fails to balance resources across the multiple objectives (cpu, mem, task count) in one solution mapping.

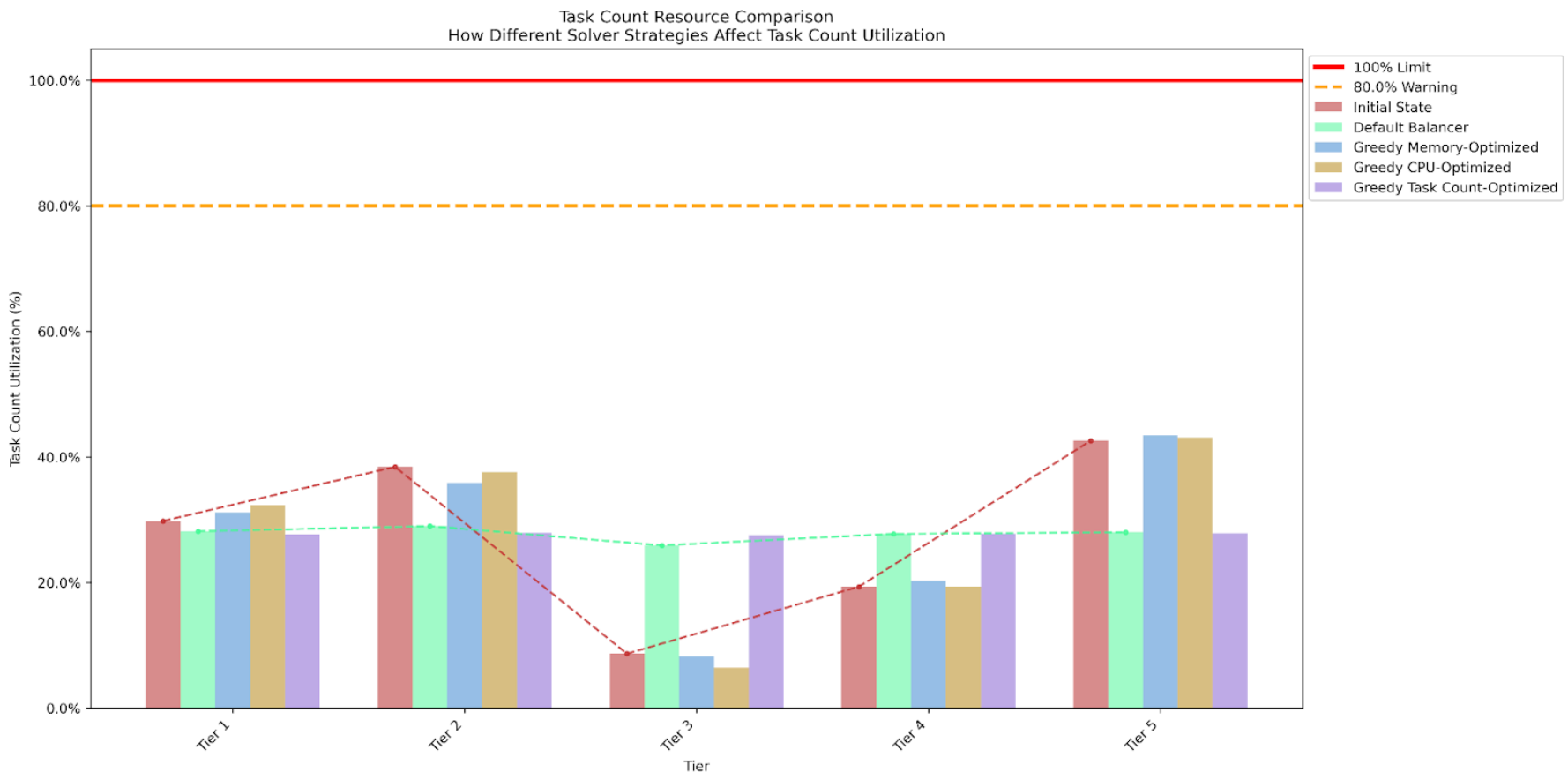

(c) In Figure 3(c) we have task count per tier on the y-axis and the respective tier bar plots in the x-axis. Unlike Figures 3(a), (b), the tiers have a default ideal task count at 80%. The percentage is relative to each tier's capacity limit. The graph represents live data collected from Meta's infrastructure. The red bar represents the initial state before the SPTLB is run. Neon-green represents SPTLB's balanced mapping for app to tier, we can see the bar's are much more comparable in height from the initial red positions. We see the same pattern from Figure 3(a), (b), where the baseline greedy solver does perform load balancing effectively enough on one resource at a time, but fails to balance resources across the multiple objectives (cpu, mem, task count) in one solution mapping.

Figure 3 (a, b, c): Each graph depicts the objective we're optimizing for cpu, mem and task count. Each graph indicates the performance of the SPTLB scheduler and the greedy scheduler variants which prioritize cpu, memory and task count respectively.

**4.2.2 Network cost across SPTLB integration variations**

In the graph of Figure 4 we see the network latency cost relations for each integration variant. We have three variants to the SPTLB scheduler.

1. No additional constraints(No_cnst): This variant is where there is no explicit attempt to make any integration between SPTLB and its lower-level solvers.
2. With additional constraints(W_cnst): This variant is where we explicitly state for each tier their regions and say that for an app to transition from one tier to the other, the tiers must share a majority of regions (>50% of regions in tier 1 must overlap with tier 2 to be considered a valid transition the SPTLB can make). This variant is stated as additional constraints for the scheduler, therefore vastly increasing its complexity but making it region aware to be compatible with the lower-level region scheduler.
3. Manually Added Constraint(Manual_cnst): This variant is meant to represent the proposed methodology of inter-communication from section 3.4. Here we manually add constraints to deter transitions that were detected in the no_cnst variant as high latency transitions and then manually add them to the solver, thus emulating the condition of being accepted or rejected by the region scheduler.

In Figure 4 we plot the p99 values from the CDF of the latency distributions for each source to destination tier mapping generated by the SPTLB scheduler during balancing. This latency distribution is then randomly sampled a 1000 times based on the number of apps selected for that particular source to destination tier combination. We then create a CDF from these values to showcase what the worst case scenario network latency would look like. The below values are approximated to the closets ms. We also showcase optimal search against local search for the solutions with varying times as specified in the x-axis. Here we see the trend that no_cnst is the worst variant to drop the network latency cost, given that it's completely un-aware of the network scheduling this makes sense. In an effort to reduce this we try no_cnst which does successfully reduce this latency, however with the additional complexity we do see the w_cnst variant more often takes more time to generate a solution than the other variants. Finally our manual_cnst variant also succeeds in dropping the worst case latency cost from the no_cnst variant. Albeit not as well as the w_cnst variant, but it does get close.

However this is only a trend for the network, since we know that the w_cnst and manual_cnst variants are much more limited in their transitions than the no_cnst variant, we need to ensure we're still successfully load balancing the applications across the tiers.

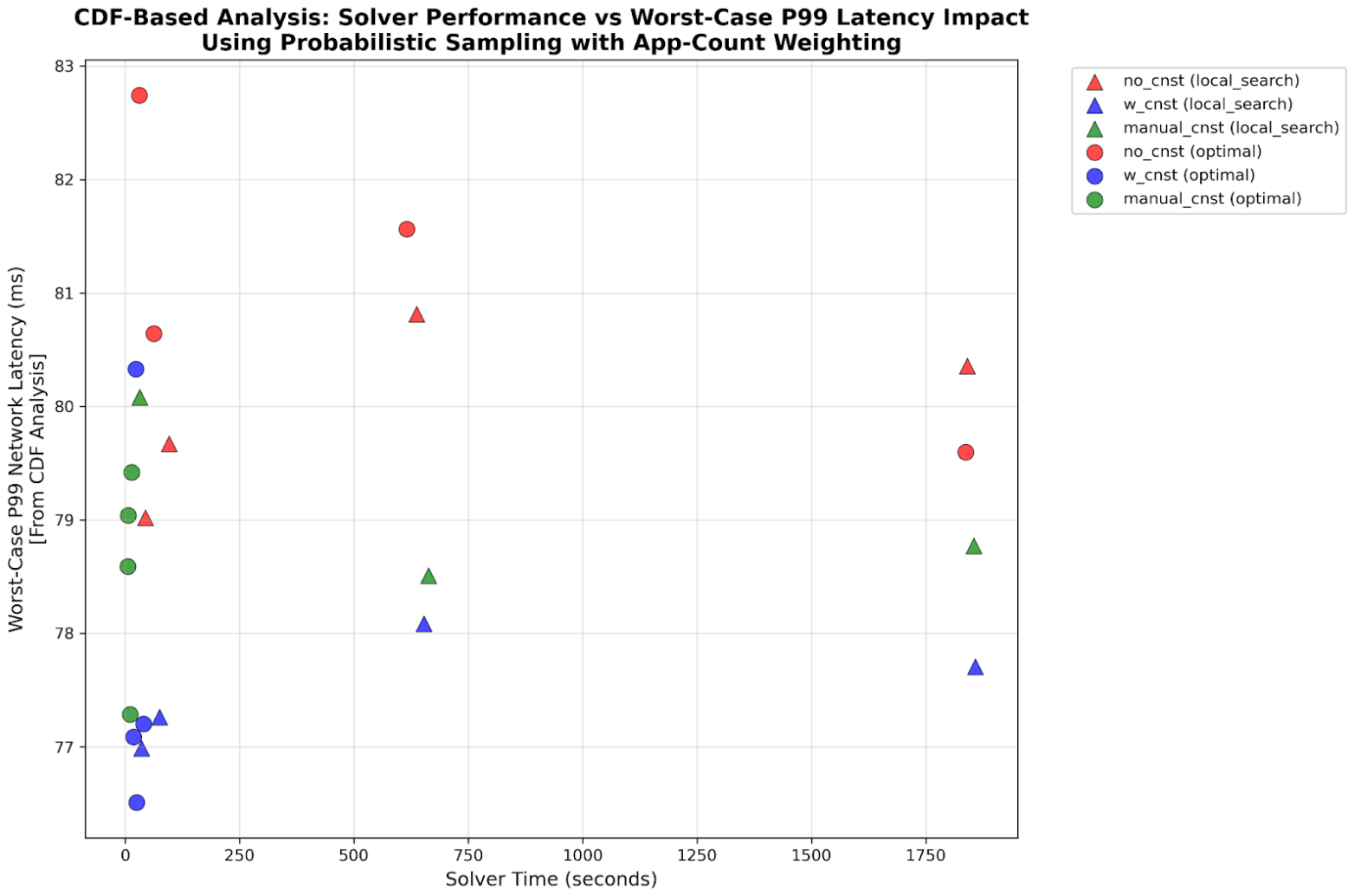

Figure 4: The graph plots the worst case (99th percentile or p99) network latency that can be incurred during an app movement between tiers. Each point represents the p99 of a CDF sampled uniformly at random across the source and destination tier's region latency table. This CDF is also dependent on the number of apps being moved, with distribution being sampled proportionally to apps moved for those tier transitions. The triangles represent solutions from the local search solver and the dots represent the solutions from the optimal search solver. The x-axis is the time taken by solver to generate a solution given a timeout of 30s, 60s, 600s (10m) and 1800s (30m). We can see that the network latency for the w_cnst variants are almost always better than the other variants. No_cnst is the worst performing solution and manual_cnst is the middle ground where it sometimes beats w_cnst and sometimes does not. However this graph is only 1-dimensionally looking at network cost, later on in Figure 5, we see the trade-off between network cost and balancing of other resources more clearly and find the pareto optimal solution, indicating an effective way to communicate between these multi-objective hierarchy of schedulers.

#### 4.2.3 Overall solution quality across SPTLB integrations

When looking at the full picture for the resource balancing we see that in Figure 5, a pareto frontier is formed. Here we have, difference to balanced state (y-axis) which is the relative difference from the ideal balanced state of each resource (50%), with the final mapping provided by the SPTLB scheduler variants, Each point represents the worst balanced resource difference, much like the reasoning for taking p99 for our network latency distribution, we also want to account for the worst case in our comparisons. This means the lower the point on the y-axis the more balanced the solution provided by SPTLB is as the difference to balance is smaller. Our x-axis represents time taken to generate a solution given a timeout of 30s, 60s, 10 minutes, 30 minutes. We see that, as hypothesised previously, in Sections 3.4 and Sections 4.2.2, w_cnst performs much worse, in both solution quality and time taken, than the others due to its increased complexity and high restrictions on app transitions even though it performs well in reducing network latency. And we see that no_cnst perform better as expected given the lack of constraints on the app transitions, but it performs poorly for network latency, however the pareto frontier is shown by the manual_cnst points, where not only do we get the best solution, we also get it in the least amount of time. This showcases the optimal way of cooperation between hierarchical, independent schedulers. The optimal searches (the dots in the graph) do not seem to consistently perform better or worse than the local searches depending on the variant and could be the result of too small of a timeout, given the complexity of the system.

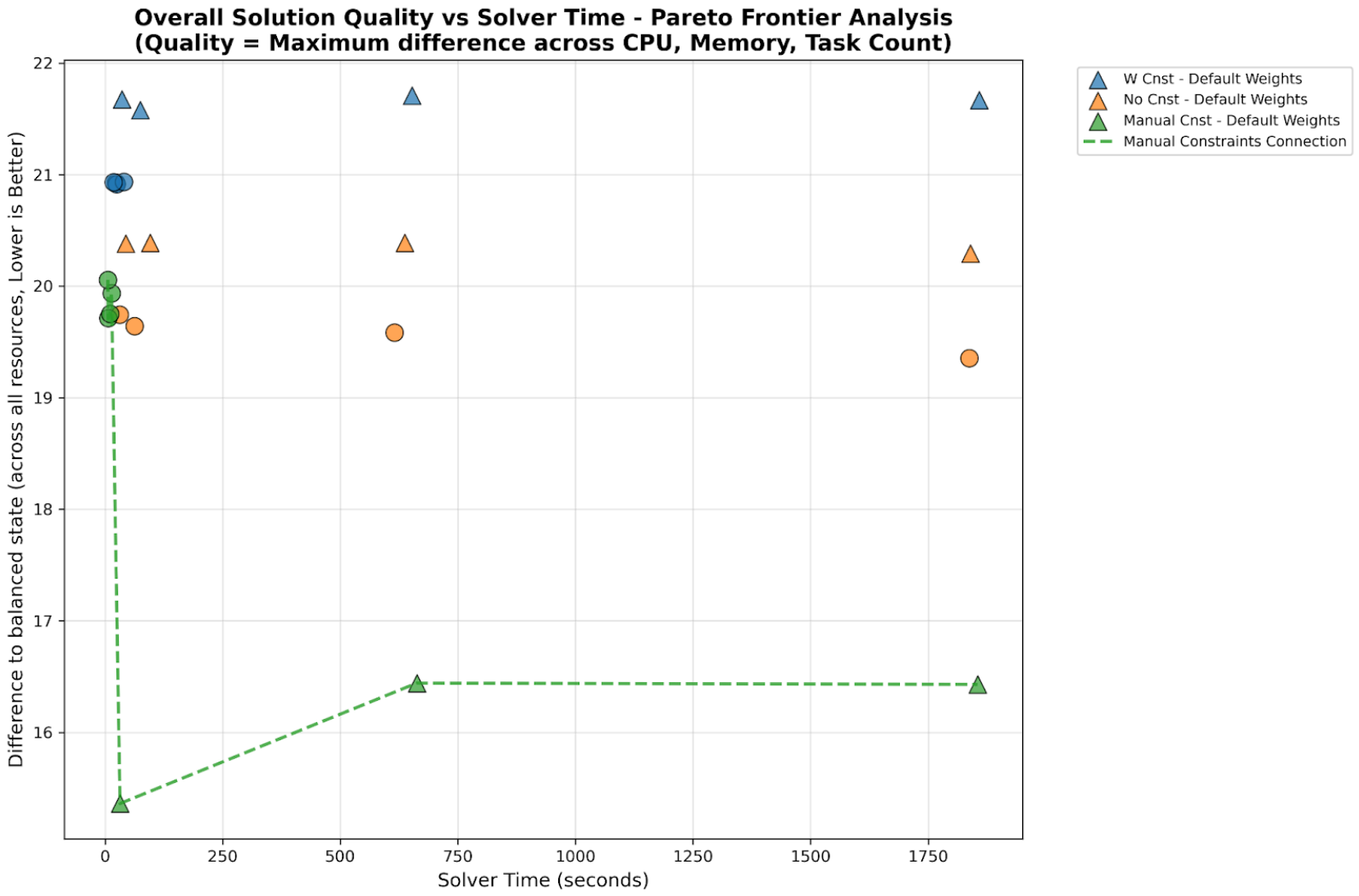

Figure 5: In this graph we plot the pareto frontier analysis of the different SPTLB variants for integration with the system of hierarchical schedulers. Here our x-axis indicates the difference to balanced state, which means the difference between the final state mapping output from the SPTLB variant and an even distribution of said resource given the initial states, in our case, 50% is the ideal balanced scenario, we take this value as the maximum difference across all resources, indicating the worst case scenario for balancing. Therefore the lower the difference, the closer we are to a balanced state. The y-axis indicates the time taken to generate the final mapping by each SPTLB variant. The lower the time taken the better. And the dashed line indicates the pareto frontier which shows to correspond to all the values from the manual_cnst variant, indicating that the ideal solver co-operation methodology is manual_cnst.

## 5. Conclusion and Future Work

From our results section we successfully showcase our contributions by:

- Establishing our SPTLB's performance on load balancing across multiple objectives (cpu, mem, task count).
- Validating its effectiveness when compared to a greedy scheduler that emulates manual app movement.
- And we also propose and empirically show our methodology of cooperation between schedulers at varying hierarchies is pareto optimal and introduces minimal added complexity and time taken to generate a solution.

Next steps in this work include validating the co-operating methodology as novel, generically applicable and effective by incorporating it into Meta's production system for the entire workflow as shown in Figure 2.